Vector Laguerre-Gauss beams with polarization-OAM entanglement in a graded-index medium

N.I. Petrov

*Scientific and Technological Center for Unique Instrumentation of Russian Academy of Sciences, Butlerova str., 15, Moscow 117342, Russia*

*Corresponding author:* petrovni@mail.ru

It is shown that the vector-vortex Laguerre-Gauss modes with polarization-orbital angular momentum (OAM) entanglement are the vector solutions of the Maxwell equations in a graded-index medium. Focusing of linearly and circularly polarized vortex light beams with nonzero azimuthal and radial indices in a cylindrical graded-index medium is investigated. The wave shape variation with distance taking into account the spin-orbit and nonparaxial effects is analyzed. Effect of long-term periodical revival of wave packets due to mode interference in a graded-index cylindrical optical waveguide is demonstrated. High efficiency transfer of a strongly focused spot through an optical waveguide over large distances takes place with a period of revival.

In a scalar approximation the light beam propagating in an inhomogeneous medium is governed by diffractive and refractive forces. However a consistent electromagnetic theory of light propagation should be originated from the vector Maxwell equations. When a vector wave field is considered, the influence of additional effective forces associated with the polarization (spin angular momentum (SAM)) and orbital angular momentum (OAM) should be taken into account [1]. It is well established that the polarization vector of linearly polarized optical beams propagating over a spiral trajectory undergoes the Rytov rotation [2, 3]. It is also of interest to consider the inverse effect, i.e. the influence of polarization on the trajectory and the width of a radiation beam. Recently, the influence of polarization on the trajectory of a light beam propagating in an inhomogeneous medium (the optical Magnus effect [4, 5] or optical spin Hall effect [6, 7]) has been predicted and studied. It was shown in [8] that spin-orbit interaction causes an asymmetry effect for depolarization of the right- and left-handed circularly polarized light propagating in a graded-index fiber. Note that besides SAM and OAM, light beams possess the radial number associated to the intensity distribution in the transverse plane [9-11].

In this paper the influence of polarization (spin) and OAM on a vortex light beam shape with non-zero radial number propagating through the cylindrically symmetric waveguide with a gradient profile of the refractive index is investigated by solving the full three-component field Maxwell equations. Remote focusing and high efficiency transfer of a focused spot in a graded-index waveguide is demonstrated. For paraxial beams, the different modes or rays turn periodically to be in phase during their propagation in graded-index optical waveguides, and periodical focusing and defocusing can be observed [12]. However, nonparaxial effects destroy this periodical focusing. Surprisingly, a restoration of an initial beam shape and a strong focusing occur at extremely long distances due to the revival effect caused by the interference between propagating modes [13]. The expressions for the trajectory and width of the beam describing the non-paraxial propagation in GRIN media were obtained earlier in [14, 15]. In [16] the revival effect in GRIN media up to second order approximation was considered. In [17] the exact analytical expressions for the trajectory and width of nonparaxial wave beams were obtained. Note that the revival effect in cylindrical GRIN media produced by non-paraxial propagation were also recently studied in [18]. However, the problem was considered in the frame of the scalar Helmholtz equation, which is not the case of nonparaxial propagation of a light beam in a cylindrical waveguide.

The Maxwell equations for the electric field $\mathbf{E}\exp(-i\nu t)$ in a general inhomogeneous medium with the dielectric constant $\varepsilon(x, y)$ reduce to [19]:

$$\begin{aligned}&\left(\nabla_{\perp}^{2}+k^{2}n^{2}\right)\mathbf{e}_{\perp}-i\beta\boldsymbol{\nabla}_{\perp}e_{z}-\boldsymbol{\nabla}_{\perp}\boldsymbol{\nabla}_{\perp}\mathbf{e}_{\perp}=\beta^{2}\mathbf{e}_{\perp}\\&\left(\nabla_{\perp}^{2}+k^{2}n^{2}\right)e_{z}+i\beta\mathbf{e}_{\perp}\boldsymbol{\nabla}_{\perp}\ln n^{2}=\beta^{2}e_{z}\end{aligned} \qquad (1)$$

where $k = 2\pi/\lambda$ is the wavenumber and $n^2(x, y)$ is the dielectric permittivity of the medium, $\beta$ is the propagation constant.

The absence of medium nonlinearity and loss is assumed below.

Consider a rotationally symmetric cylindrical waveguide with a parabolic distribution of the refractive index:

$$n^2 = n_0^2 - \omega^2 r^2 \ , \ 0 \le r \le a \, , \qquad (2)$$

where $n_0$ is the refractive index on the waveguide axis, $\omega$ is the gradient parameter, $r = \sqrt{x^2 + y^2}$, $a$ is the radius of the waveguide.

The equation (1) may be rewritten in terms of annihilation and creation operators in cylindrical coordinates, i.e. the unperturbed Hamiltonian $\widehat{H}_0 = \left(\omega / k n_0^2\right)\left(\widehat{A}_1^{+}\widehat{A}_1 + \widehat{A}_2^{+}\widehat{A}_2 + 1\right)\widehat{I}$ [19]. The representation of the Hamiltonian via the operators will allow us to calculate the matrix elements analytically with the help of pure algebraic procedure. The solution of the unperturbed equation is described by radially symmetric Laguerre-Gauss functions $\psi_{vl}(r,\varphi) = |v,l\rangle$:

$$\psi_{vl}(r,\varphi) = \left(\frac{k\omega}{\pi}\right)^{1/2} \left[\frac{p!}{(p+l)!}\right]^{1/2} \left(k\omega r^2\right)^{l/2} \exp\left(-\frac{k\omega r^2}{2}\right) L_p^l\left(k\omega r^2\right) \exp(il\varphi), \qquad (3)$$

where $v = 2p + l$ is the principal quantum number, $p$ and $l$ are the radial and azimuthal indices, accordingly, and $l = v, v-2, v-4, \ldots 1$ or 0, $\omega = 2/(kw_0^2)$, $w_0$ is the radius of the fundamental mode. The numbers $v$ and $l$ express the eigenvalues of the unperturbed Hamiltonian $\hat{H}_0 |v,l\rangle = \left(\omega / kn_0^2\right)(v+1)|v,l\rangle$, and eigenvalues $L = l/k$ of the angular momentum operator $\hat{L}_z |v,l\rangle = (l/k)|v,l\rangle$.

It was shown in [19] that the hybrid wave functions consisting of transverse and longitudinal components are the solutions of the equation (1):

$$\mathbf{\Psi}(r,\varphi,0) = \left| \begin{matrix} |vl\rangle \\ i\sigma|vl\rangle \\ e_z \end{matrix} \right\rangle, \qquad (4)$$

where $\sigma = +1$ and $\sigma = -1$ correspond to right-handed and left-handed circularly polarized beams, accordingly, and $\sigma = 0$ corresponds to the linear polarization.

There is no mode conversion at propagation if the incident beam is expressed by (4). Note that the hybrid wave function (4) cannot be factorized into the product of spin and orbital parts since the mixing of OAM and SAM exists. Thus, the modal solutions of the Maxwell equations in a GRIN media are the hybrid vector Laguerre-Gauss modes with the spin-orbit entanglement. The longitudinal field component can be expressed through the transverse field components, i.e. $|e_z\rangle = (i/kn_0)\nabla_\perp \mathbf{e}_\perp$.

The propagation constant correct to first-order nonparaxial term is given by [20]:

$$\beta_{vl\sigma} = kn_0 \left\{ 1 - \eta(v+1) - \frac{\eta^2}{32}\left[ 11(v+1)^2 - j^2 - 2j\sigma \right] \right\}, \tag{5}$$

where $\eta = \omega / kn_0^2$, $j = l + \sigma$.

Consider the incident vector vortex beams with right-circular and left-circular polarizations, accordingly: $\langle \mathbf{\Psi}_0^+ | = \left( \langle v'l |, -i\langle v'l |, e_z \right)$ and $\langle \mathbf{\Psi}_0^- | = \left( \langle v'l |, i\langle v'l |, e_z \right)$, where $| v'l \rangle$ is given by (3), and $\omega' = 2/(ka_0^2)$, $a_0$ is the radius of a beam which is different from the radius of the fundamental mode of the medium $w_0 = \sqrt{2/(k\omega)}$.

The arbitrary incident beam may be expanded into modal solutions, so the evolution of a beam in the medium (2) can be represented as

$$\mathbf{\Psi}(r,\varphi,z) = \sum_{vl\sigma} a_{vl\sigma} \left| \begin{matrix} |vl\rangle \\ i\sigma |vl\rangle \\ (i/kn_0)\vec{\nabla}_\perp (\vec{x} + i\sigma\vec{y}) |vl\rangle \end{matrix} \right\rangle \exp\left(i\beta_{vl\sigma} z\right), \tag{6}$$

where $a_{vl\sigma}$ are the coupling coefficients.

Below only the propagating modes are considered, i.e. all propagating power is located inside the waveguide with the radius $a$, the evanescent waves do not reach the far-field zone.

If the incident beam is described by the Laguerre-Gauss function $\mathbf{\Psi}^*_{v'l'\sigma} = \left(1/\sqrt{2}\right)\left\langle \langle v'l' |, -i\sigma\langle v'l' |, e_z^* \right|$, the coupling coefficients $a_{vl\sigma}$ can be calculated analytically:

$$\langle vl\sigma | v'l\sigma \rangle = \left( \frac{2\sqrt{\omega\omega'}}{\omega + \omega'} \right)^{l+1} \left( \frac{\omega' - \omega}{\omega' + \omega} \right)^{p-p'} \left( \frac{p'!(p+l)!}{(p'+l)!\,p!} \right)^{1/2} P_{p'}^{[p-p',l]}(z), \tag{7}$$

where $z = 1 - 2\left( \frac{\omega' - \omega}{\omega' + \omega} \right)^2$, $P_{p'}^{[p-p',l]}(z)$ are the Jacobi polynomials, $\omega' = 2/ka_0^2$, $\omega = 2/kw_0^2$.

The wave shape variations with distance are determined by the functions $I_{\perp}(r,\varphi,z)=\left|\boldsymbol{\psi}_{\perp}(r,\varphi,z)\right|^2$ and $I_z(r,\varphi,z)=\left|e_z(r,\varphi,z)\right|^2$. Fig. 1 shows the intensity profiles of the linearly polarized beams with the radial number $p'=1$ and different OAM in the focal plane. The medium (2) with the gradient parameter $\omega=7\times10^{-3}\,\mu m^{-1}$ and the refractive index $n_0=1.5$ is considered. These parameters are reasonable for conventional graded-index optical fibers. Here and below the beams with wavelength $\lambda=0.63\mu m$ are considered. The numerical aperture is determined by $NA=a\omega=n_0\sqrt{2\Delta}$, where $a$ is the radius of the waveguide and $\Delta\approx(n_0-n(a))/n_0$. The initial beam width or the full width at half maximum (FWHM) is $a_0=45\mu m$. It is seen that the intensity distributions depend on the SAM and OAM of the incident beam. For $l=0$ the focused spot in the longitudinal field component is splitting into two equal parts (Fig.1b). There is an asymmetry between the longitudinal field component intensity distributions for the incident beams with opposite OAM (Fig.1d and Fig.1f). Note, that there is no such asymmetry for the transverse field components.

The wave shapes of circularly polarized beams at various distances were also calculated. Fig. 2 displays the intensity profiles of the circularly polarized beams with different OAM in a waveguide with $\omega=7\cdot10^{-3}\mu m$ and NA=0.5 at various distances. It is seen that the beams with antiparallel OAM and SAM can achieve tighter focal spots than those for which the signs of the helicity and the orbital angular momentum are the same. Similar result was obtained from the uncertainty-type relations between focal spot size and angular spread [21]. High efficiency transfer of a focused spot through optical waveguide over large distances with a period of revival takes place.

The width of the incident beam oscillates with a period of $L_0=\pi n_0/\omega$ and these oscillations relax to the static value determined by the width of an incident beam. Focusing of a beam takes place at a distance of $L_0^f=\pi n_0/(2\omega)$. The long-term periodic revivals of the initial width and focusing occur with a period close to $L_{rev} \cong \pi m/(2\gamma)$, where $m$ = 1, 2 …, $\gamma = \omega^2/(kn_0^3)$. Note that $L_0 << L_{rev}$. Note that the focal planes are located at different distances for the beams with different quantum numbers.

Consider the propagation of a strongly focused Gaussian beam in a medium (2). Note that only the propagating modes reach the far-field zone. For the beam with $a_0 \geq \lambda/(2NA)$ all incident power is in the propagating modes and the periodical revivals of the initial field intensity distribution occur at extremely long distances. In Fig.3 the intensity distributions of the propagating fields in the transverse initial plane $z = 0$ and at long-term revival distances z=38.47mm and z=38.45mm, accordingly, in a nanofiber with $NA = 1.0$ for different azimuthal and radial indices are shown. The power contained in the propagating modes amounts to $P_\perp \approx 32.4\%$ and $P_z \approx 1.8\%$ of the total beam power if $p = 1$ and $l = 0$ and $P_\perp \approx 60\%$ and $P_z \approx 7.3\%$ if $p = 0$ and $l = 1$. Note, that the spots and doughnut rings with a FWHM smaller than $\lambda/(2NA)$ can be transferred with the help of propagating modes (Fig.3). It is followed from the calculations that the periodical revivals of the initial field intensity distribution occur at extremely long distances and the high efficiency transfer of the subwavelength spot through optical waveguide over large distances takes place with a period of revival.

In conclusion, propagation of vector vortex beams in an inhomogeneous medium is analysed by solving three-component field Maxwell equations. The polarization-dependent properties of the electric field intensity profiles in the focal plane are examined for the beams with OAM and SAM. The asymmetry effects manifesting in different intensity distributions in the focal plane for opposite handedness of vorticity and/or polarization are demonstrated. It is shown that the beams with antiparallel OAM and SAM can be focused into tighter spots than those for which these angular momentum are parallel. The influence of azimuthal and radial mode indices on focusing spot is investigated. The fundamental effect of collapse and revival of wave packets at the propagation in rotationally symmetric waveguide is examined. The long-term periodical revival of a focused incident beam profile taking into account the spin-orbit and nonparaxial effects is demonstrated. Due to this effect the remote subwavelength focusing of a light beam in an optical fiber can be achieved without the use of the evanescent waves. Vector modal solutions exhibiting entanglement between spin (polarization) and OAM (wavefront vorticity) may be useful for classical implementations of quantum communication and computational tasks [22].

Figure captions:

**Fig.1.** Intensity profiles of the transverse electric field component (left column) and the longitudinal electric field component (right column) for the linearly polarized incident beam with nonzero radial number in the focal planes $z_f = 331\mu m$ (a-d) and $z_f = 333\mu m$ (e-h): (a, b) $l = 0$; (c, d) $l = 0$ - 3D intensity patterns; (e, f) $l = 1$; (g, h) $l = -1$. $NA = 0.5$.

**Fig.2.** Intensity profiles of the transverse electric field component (left column) and the longitudinal electric field component (right column) for the circularly polarized incident beams in the focal planes (a, b), $z$ = 78.48 mm (d-f) $z$ = 78.46 mm. (a, b) $l = 0$, $\sigma = 1$; (c, d) $l = 1$, $\sigma = 1$; (e, f) $l = 1$, $\sigma = -1$.

**Fig.3.** Long-term transfer of a subwavelength focused spot in a nanofiber with $\omega = 0.1\mu m^{-1}$. Intensity profiles of the transverse electric field components (left column) and the longitudinal electric field components (right column) in the initial plane $z = 0$ (a, b, e, f) and at long-term revival distances z= 38.47mm (c, d) and z=38.45mm (g, h): (a-d) $p = 0, l = 1, \sigma = -1$; (e-h) $p = 1, l = 0, \sigma = 1$.

Fig.1

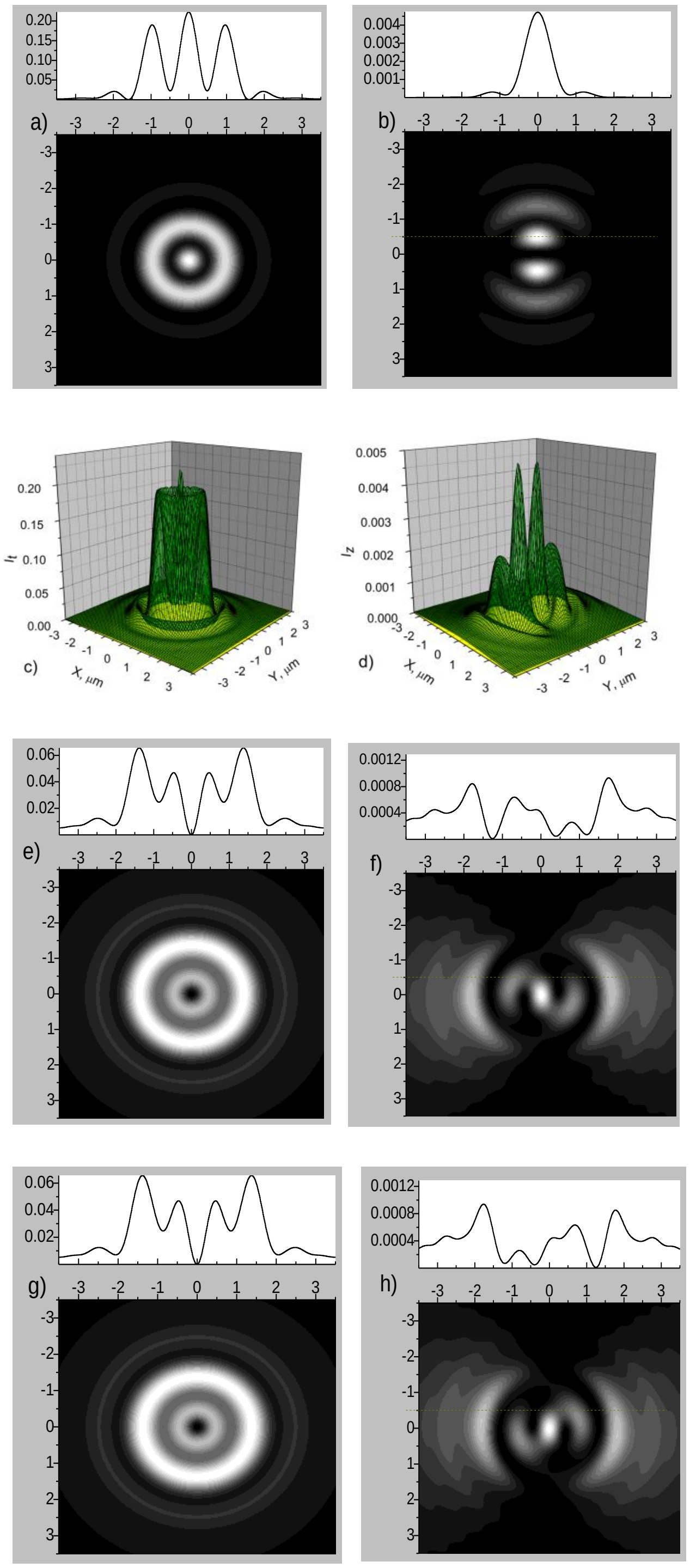

Fig.2

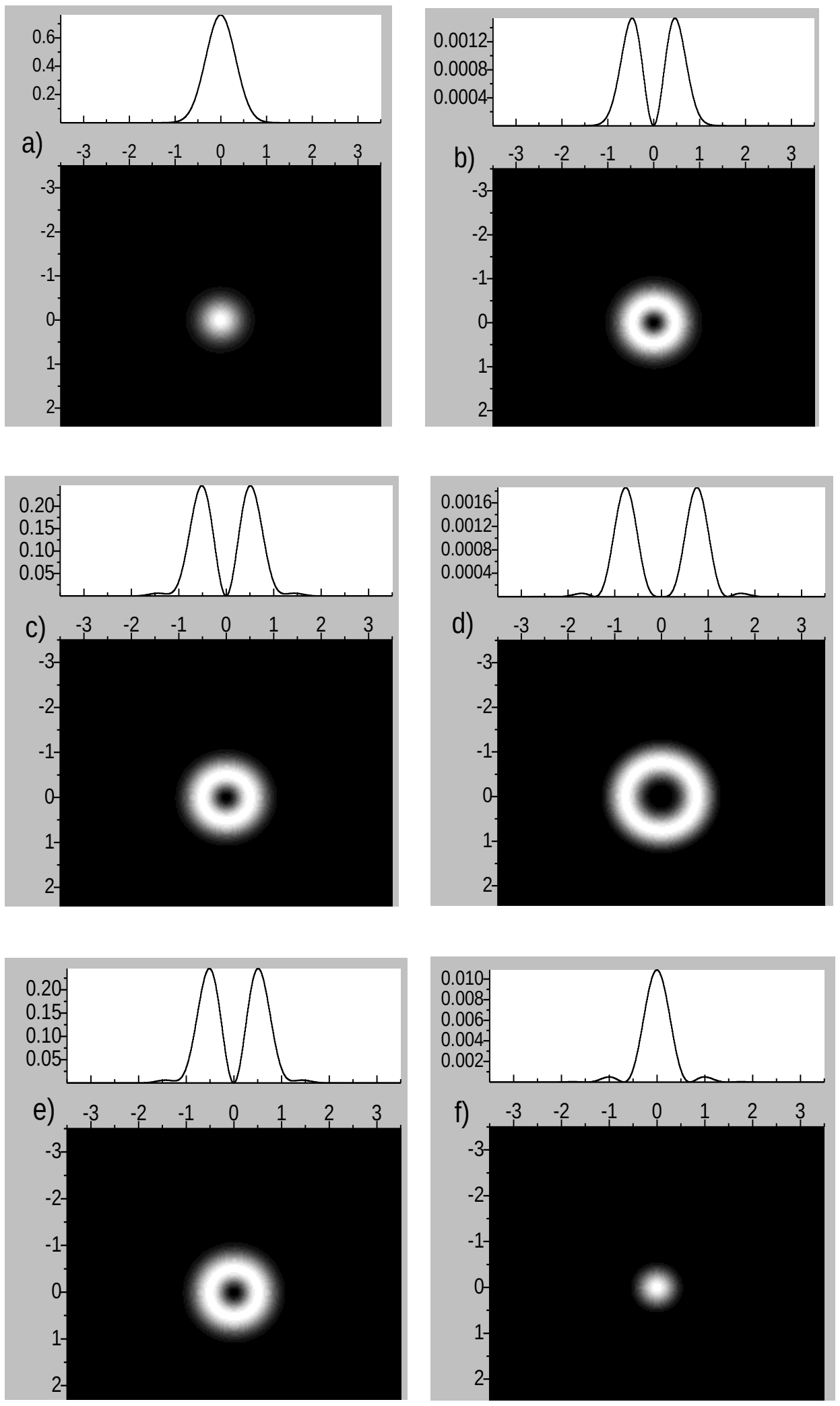

Fig.3

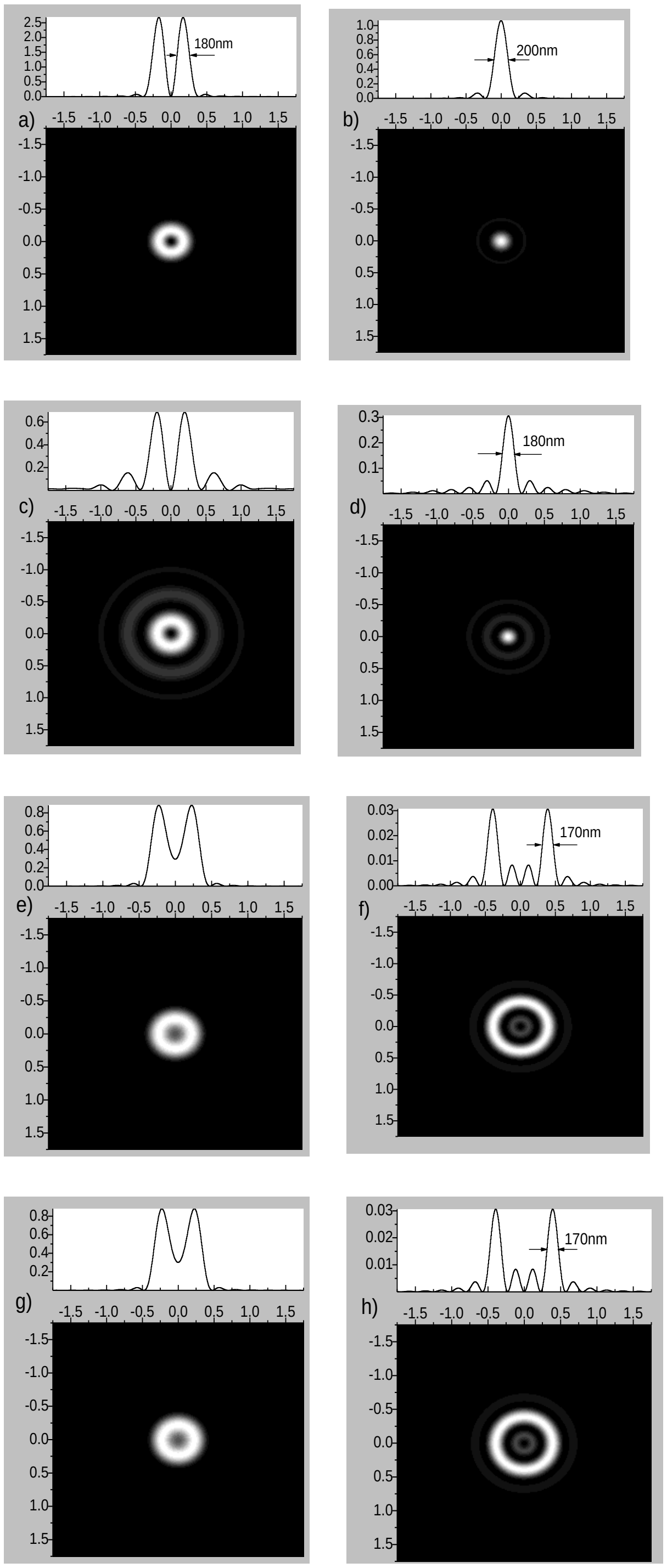